%% file: coopis03.tex
\documentclass[runningheads]{llncs}
\usepackage{latexsym}

\input new_fig.tex

\def\punto{\hspace*{\fill}\Box}

\date{}
\begin{document}

\title{``Almost automatic'' and semantic integration of XML Schemas at various ``severity'' levels}

\titlerunning{``Almost automatic'' and semantic integration...}


\author{Pasquale De Meo\inst{1} \and Giovanni Quattrone\inst{1} \and \\
Giorgio Terracina\inst{2} \and Domenico Ursino\inst{1}}
\authorrunning{Pasquale De Meo \and Giovanni Quattrone \and Giorgio Terracina \and Domenico Ursino}
\tocauthor{Pasquale De Meo (DIMET, Universit\`a Mediterranea di Reggio Calabria), Giovanni
Quattrone (DIMET, Universit\`a Mediterranea di Reggio Calabria), Giorgio Terracina (Dip. di
Matematica, Universit\`a della Calabria), Domenico Ursino (DIMET, Universit\`a Mediterranea di
Reggio Calabria)} \institute{DIMET, Universit\`a Mediterranea di Reggio Calabria, Via Graziella,
Localit\`a Feo di Vito, 89060 Reggio Calabria, Italy, \and Dipartimento di Matematica, Universit\`a
della Calabria, Via Pietro Bucci,
87036 Rende (CS), Italy\\
\email{demeo@ing.unirc.it, quattrone@ing.unirc.it, \\ terracina@mat.unical.it, ursino@unirc.it}}

\maketitle

\begin{abstract}

This paper presents a novel approach for the integration of a set of XML Schemas. The proposed
approach is specialized for XML, is almost automatic, semantic and ``light''. As a further,
original, peculiarity, it is parametric w.r.t. a ``severity'' level against which the integration
task is performed. The paper describes the approach in all details, illustrates various theoretical
results, presents the experiments we have performed for testing it and, finally, compares it with
various related approaches already proposed in the literature.

\end{abstract}

\section{Introduction}
\label{Introduction}

The Web is presently playing a key role for both the publication and the exchange of information
among organizations. As a matter of fact, it is becoming the reference infrastructure for most of
the applications conceived to handle interoperability among partners.

In order to make Web activities easier, W3C (World Wide Web Consortium) proposed XML (eXtensible
Markup Language) as a new standard information exchange language that unifies representation
capabilities, typical of HTML, and data management features, typical of classical DBMS.

The twofold nature of XML allowed it to gain a great success and, presently, most of the new
documents published on the Web are written in XML. However, from the data management point of view,
XML documents alone have limited and primitive capabilities. In order to improve these
capabilities, in such a way to make them similar to those typical of classical DBMS, W3C proposed
to associate XML Schemas with XML documents. An XML Schema can be considered as a sort of catalogue
of the information typologies that can be found in the corresponding XML documents; from another
point of view, an XML Schema defines a reference context for the corresponding XML documents.

Certainly, XML exploitation is a key step for improving the interoperability of Web information
sources; however, that alone is not enough to completely fulfill such a task. Indeed, the
heterogeneity of data exchanged over the Web regards not only their formats but also their
semantics. The use of XML allows format heterogeneity to be faced; the exploitation of XML Schemas
allows the definition of a reference context for exchanged data and is a first step for handling
semantic diversities; however, for a complete and satisfactory management of these last, an
integration activity is necessary.

This paper provides a contribution in this setting and proposes an approach for the integration of a set of
XML Schemas. Our approach behaves as follows: first it determines interscheme properties
\cite{CaDeDe01,FaKrNe91,MaBeRa01,RaBe01,Ursino-TKDE1}, i.e., terminological and structural relationships
holding among attributes and elements belonging to involved XML Schemas. After this, some of the derived
properties are exploited for modifying involved Schemas in order to make them structurally and semantically
uniform. The modified Schemas are, finally, integrated for obtaining the global Schema.

Let us now examine the peculiarities of our approach in more detail. First, it has been
specifically conceived for operating on XML sources. In this sense it differs from many other
approaches already presented in the literature which integrate information sources having different
formats and structure degrees (e.g., relational databases, XML documents, object-oriented sources
and so on). Generally, such approaches translate all involved information sources into a common
representation and, then, carry out the integration activity. On the contrary, our approach is
specialized for integrating XML Schemas. With regard to this, it is worth pointing out that: {\em
(i)} the integration of XML Schemas will play a more and more relevant role in the future; {\em
(ii)} the exploitation of generic approaches designed to operate on information sources with
different formats, for performing the integration of a set of XML Schemas (i.e., a set of sources
having the same format), is unnecessarily expensive and inefficient. Indeed, it would require the
translation of involved XML Schemas in another format and the translation of the integrated source
from such a format back to XML.

Our approach is almost automatic; in this sense it follows the present trend relative to
integration techniques. Indeed, owing to the enormous increase of the number of available
information sources, all integration approaches proposed in the last years are semi-automatic;
generally, they require the human intervention for both a pre-processing phase and the validation
of obtained results. The overwhelming amount of sources available on the Web leads each integration
task to operate on a great number of sources; this requires a further effort for conceiving more
automatic approaches. The approach we are proposing here provides a contribution in this setting
since it is almost automatic and requires the user intervention only for validating obtained
results.

Our approach is ``light''; with regard to this we observe that most of the existing approaches are
quite complex, based on a variety of thresholds, weights, parameters and so on; they are very
precise but difficult to be applied and fine tuned when involved sources are numerous, complex and
belonging to heterogeneous contexts. Our approach does not exploit any threshold or weight; as a
consequence, it is simple and light, since it does not need a tuning activity.

Our approach is semantic in that it follows the general trend to take into account the semantics of
concepts belonging to involved information sources during the integration task
\cite{CaDeDe01,DoDoHa01,FaKrNe91,MaBeRa01}. Given two concepts belonging to different information
sources, one of the most common way for determining their semantics consists of examining their
neighborhoods since the concepts and the relationships which they are involved in contribute to
define their meaning. As a consequence, two concepts, belonging to different information sources,
are considered semantically similar and are merged in the integrated source if their neighborhoods
are similar.

We argue that all the peculiarities we have examined above are extremely important for a novel approach
devoted to integrate XML Schemas. However, the approach we are proposing here is characterized by a further
feature that, in our opinion, is extremely innovative and promising; more specifically, it allows the choice
of the ``severity level'' against which the integration task is performed. Such a feature derives from the
consideration that applications and scenarios possibly benefiting of an integration task on the Web are
numerous and extremely various. In some situations (e.g., in Public Administrations, Finance and so on) the
integration process must be very severe in that two concepts must be merged only if they are strongly
similar; in such a case a high severity degree is required. In other situations (e.g., tourist Web pages) the
integration task can be looser and can decide to merge two concepts having some similarities but presenting
also some differences. At the beginning of the integration activity our approach asks the user to specify the
desired ``severity'' degree; this is the only information required to her/him until the end of the
integration task, when she/he has to validate obtained results. It is worth pointing out that, to the best of
our knowledge, no approaches handling the information source integration at various ``severity'' levels have
been previously presented in the literature. Interestingly enough, a classical approach can be seen as a
particular case of that presented in this paper in which a severity level is fixed and all concept merges are
performed w.r.t. this level.

\section{Neighborhood Construction}
\label{Neighborhood-Construction}

In this section we formally introduce the concept of neighborhood of an element or an attribute of
an XML Schema. As pointed out in the Introduction, this concept plays a key role in the various
algorithms which our approach consists of. Preliminarily we introduce the concept of x-component
which allows both elements and attributes of an XML document to be uniformly handled.

\begin{definition}
\label{x-component} {\em Let $S$ be an XML Schema; an {\em x-component} of $S$ is either an element
or an attribute of $S$. $\punto$ }
\end{definition}

\noindent An x-component is characterized by its name, its typology (indicating if it is either a
complex element or a simple element or an attribute) and its data type.

\begin{definition}
\label{intens} {\em Let $S$ be an XML Schema; the set of its x-components is denoted as
$XCompSet(S)$. $\punto$ }
\end{definition}

\noindent We introduce now some boolean functions that allow to determine the strength of the
relationship existing between two x-components $x_S$ and $x_T$ of an XML Schema $S$. They will be
exploited for deriving interscheme properties and, ultimately, for integrating XML Schemas. The
functions are:

\begin{itemize}

\vspace*{-0.3cm}

\item $veryclose(x_S,x_T)$, that returns {\em true} if and only if: {\em (i)} $x_T=x_S$, or {\em (ii)} $x_T$
is an attribute of $x_S$, or {\em (iii)} $x_T$ is a simple sub-element of $x_S$;

\item $close(x_S,x_T)$, that returns {\em true} if and only if {\em (i)} $x_T$ is a complex sub-element of
$x_S$, or {\em (ii)} $x_T$ is an element of $S$ and $x_S$ has an $IDREF$ or an $IDREFS$ attribute referring
$x_T$;

\item $near(x_S,x_T)$, that returns {\em true} if and only if either $veryclose(x_S,x_T)=true$ or
$close(x_S,x_T)=true$; in all the other cases it returns {\em false};

\item $reachable(x_S,x_T)$, that returns {\em true} if and only if there exists a sequence of {\em
distinct} x-components $x_1, x_2, \ldots, x_n$ such that $x_S = x_1, near(x_1, x_2) = near(x_2,
x_3) = \ldots = near(x_{n-1},x_n) = true, x_n=x_T$. $\punto$

\end{itemize}

\vspace*{-0.3cm}

\noindent We are now able to compute the connection cost from $x_S$ to $x_T$.

\begin{definition}
\label{Connection-Cost} {\em

Let $S$ be an XML Schema and let $x_S$ and $x_T$ be two x-components of $S$. The Connection Cost
from $x_S$ to $x_T$, denoted by $CC(x_S,x_T)$, is defined as:

\begin{quote}

\vspace*{-0.3cm}

$CC(x_S,x_T) =  \left\{
\begin{array}{ll}
\mbox{$0$} & \mbox{if $veryclose(x_S,x_T)=true$} \\
\mbox{$1$} & \mbox{if $close(x_S,x_T)=true$} \\
\mbox{$\cal{C_{ST}}$} & \mbox{if $reachable(x_S,x_T)=true$ and $near(x_S,x_T)=false$} \\
\mbox{$\infty$} & \mbox{if $reachable(x_S,x_T)=false$} \\
\end{array}
\right. $

\vspace*{-0.3cm}

\end{quote}

\noindent where ${\cal C}_{ST}=min_{{x_A}}\ (CC(x_S,x_A)+CC(x_A,x_T))$ for each $x_A$ such that \\
$reachable(x_S,x_A)=reachable(x_A,x_T)=true$.

}
\end{definition}

\noindent We are now provided with all tools necessary to define the concept of neighborhood of an
x-component.

\begin{definition}
\label{neighborhood} {\em

Let $S$ be an XML Schema and let $x_S$ be an x-component of $S$. The $j^{th}$ neighborhood of $x_S$
is defined as:

\begin{center}

\vspace*{-0.3cm}

$neighborhood(x_S,j) = \{ x_T | \ x_T \in XCompSet(S), CC(x_S,x_T) \leq j \}$

\vspace*{-0.3cm}

\end{center}
\vspace*{-0.8cm}$\punto$ }
\end{definition}

\noindent The construction of all neighborhoods can be easily carried out with the support of the
data structure introduced in the next definition.

\begin{definition}
\label{XS-Graph}

{\em Let $D$ be an XML document and let $S$ be the corresponding XML Schema. The {\em XS-Graph}
relative to $S$ and $D$ is an oriented labeled graph defined as $XG(S,D) = \langle N(S), A(S,D)
\rangle$. Here, $N(S)$ is the set of nodes of $XG(S,D)$; there is a node in $XG(S,D)$ for each
x-component of $S$. $A(S,D)$ is the set of arcs of $XG(S,D)$; there is an arc $\langle N_S, N_T,
f_{ST} \rangle$ in $XG(S,D)$ for each pair $(x_S, x_T)$ such that $near(x_S, x_T)=true$; in
particular, $N_S$ (resp., $N_T$) is the node of $XG(S,D)$ corresponding to $x_S$ (resp., $x_T$) and
$f_{ST} = CC(x_S, x_T)$. }$\punto$

\end{definition}

\noindent The following proposition measures the computational complexity of the construction of
$XG(S,D)$.

\begin{proposition}
\label{XS-Graph-Complexity}

{\em Let $D$ be an XML document and let $S$ be the corresponding XML Schema. Let $n$ be the number of
x-components of $S$ and let $N_{inst}$ be the number of instances of $D$. The worst case time complexity for
constructing $XG(S,D)$ from $S$ and $D$ is $O(max\{n,N_{inst}^2\})$. } $\punto$

\end{proposition}

With regard to this result we observe that, in an XML document, in order to determine the element
which an IDREFS attribute refers to, it is necessary to examine the document, since neither the DTD
nor the XML Schema provide such an information. As a consequence, the dependency of the
computational complexity from $N_{inst}$ cannot be avoided. However, we point out that the
quadratic dependency from $N_{inst}$ is mainly a theoretical result; indeed, it derives from the
consideration that each IDREFS attribute could refer to $N_{inst}$ components. Actually, in real
situations, each IDREFS attribute refers to a very limited number of instances; as a consequence,
the dependency of the computational complexity from $N_{inst}$ is generally linear.

The next theorem determines the worst case time complexity for computing all neighborhoods of all
x-components of an XML Schema $S$.

\begin{theorem}
{\em Let $XG(S,D)$ be the XS-Graph associated with an XML document $D$ and an XML Schema $S$ and let $n$ be
the number of x-components of $S$. The worst case time complexity for computing all neighborhoods of all
x-components of $S$ is $O(n^3)$. } $\punto$
\end{theorem}

\begin{example}
\label{neighborhood-example}

Consider the XML Schema $S_1$, shown in Figure \ref{S1}, representing a shop. Here {\em customer}
is an x-component and its typology is ``complex element'' since it is an element declared with a
``complex type''. Analogously {\em SSN} is an x-component, its typology is ``attribute'' and its
data type is ``string''. All the other x-components of $S_1$, the corresponding typologies and data
types can be determined similarly.

In $S_1$, $veryclose(customer, firstName)=true$ because {\em firstName} is a simple sub-element of {\em
customer}; analogously $veryclose(customer, SSN)=true$ and $close(customer,musicAcquirement)=true$. As for
neighborhoods, we have that:

\begin{figure} [t]
{\tiny
\begin{minipage} [t] {6cm}
\begin{verbatim}

<?xml version="1.0" encoding="UTF-8"?> <xs:schema
xmlns:xs="http://www.w3.org/2001/XMLSchema">
    <!-- Definition of attributes -->
    <xs:attribute name="SSN" type="xs:string"/>
    <xs:attribute name="code" type="xs:ID"/>
    <xs:attribute name="acquiredBooks" type="xs:IDREFS"/>
    <xs:attribute name="acquiredMusics" type="xs:IDREFS"/>
    <xs:attribute name="acquirementDate" type="xs:date"/>
    <!-- Definition of simple elements -->
    <xs:element name="firstName" type="xs:string"/>
    <xs:element name="lastName" type="xs:string"/>
    <xs:element name="address" type="xs:string"/>
    <xs:element name="gender" type="xs:string"/>
    <xs:element name="birthDate" type="xs:date"/>
    <xs:element name="profession" type="xs:string"/>
    <xs:element name="artist" type="xs:string"/>
    <xs:element name="author" type="xs:string"/>
    <xs:element name="title" type="xs:string"/>
    <xs:element name="pubYear" type="xs:integer"/>
    <xs:element name="publisher" type="xs:string"/>
    <xs:element name="genre" type="xs:string"/>
    <xs:element name="support" type="xs:string"/>
    <!-- Definition of complex elements -->
    <xs:element name="bookAcquirement">
        <xs:complexType>
            <xs:attribute ref="acquirementDate"/>
            <xs:attribute ref="acquiredBooks"/>
        </xs:complexType>
    </xs:element>
    <xs:element name="musicAcquirement">
        <xs:complexType>
            <xs:attribute ref="acquirementDate"/>
            <xs:attribute ref="acquiredMusics"/>
        </xs:complexType>
    </xs:element>
    <xs:element name="customer">
        <xs:complexType>
            <xs:sequence>
                <xs:element ref="firstName"/>
                <xs:element ref="lastName"/>
                <xs:element ref="address"/>
                <xs:element ref="gender"/>
                <xs:element ref="birthDate"/>
                <xs:element ref="profession"/>

\end{verbatim}
\end{minipage}$ $
\begin{minipage}[t]{6cm}
\begin{verbatim}

                <xs:element ref="bookAcquirement"
                    minOccurs="0" maxOccurs="unbounded"/>
                <xs:element ref="musicAcquirement"
                    minOccurs="0" maxOccurs="unbounded"/>
            </xs:sequence>
            <xs:attribute ref="SSN" use="required"/>
        </xs:complexType>
    </xs:element>
    <xs:element name="music">
        <xs:complexType>
            <xs:sequence>
                <xs:element ref="artist" maxOccurs="unbounded"/>
                <xs:element ref="title"/>
                <xs:element ref="pubYear"/>
                <xs:element ref="genre"/>
                <xs:element ref="support"/>
            </xs:sequence>
            <xs:attribute ref="code" use="required"/>
        </xs:complexType>
    </xs:element>
    <xs:element name="book">
        <xs:complexType>
            <xs:sequence>
                <xs:element ref="author" maxOccurs="unbounded"/>
                <xs:element ref="title"/>
                <xs:element ref="publisher"/>
                <xs:element ref="pubYear"/>
                <xs:element ref="genre"/>
            </xs:sequence>
            <xs:attribute ref="code" use="required"/>
        </xs:complexType>
    </xs:element>
    <!-- Definition of root element -->
    <xs:element name="shop">
        <xs:complexType>
            <xs:sequence>
                <xs:element ref="customer" maxOccurs="unbounded"/>
                <xs:element ref="music" maxOccurs="unbounded"/>
                <xs:element ref="book" maxOccurs="unbounded"/>
            </xs:sequence>
        </xs:complexType>
    </xs:element>
</xs:schema>
\end{verbatim}
\end{minipage}
} \caption{The XML Schema $S_1$} \label{S1}
\end{figure}

\vspace*{0.2cm}

{\em neighborhood(customer,0) = \{customer, SSN, firstName, lastName, address,

\hfill{gender, birthDate, profession\}}}

\noindent All the other neighborhoods can be determined similarly. $\punto$
\end{example}

\section{Extraction of interscheme properties}
\label{Interscheme-Properties}

In this section we illustrate an approach for computing interscheme properties among x-components
belonging to different XML Schemas. As pointed out in the Introduction, their knowledge is crucial
for the integration task. The interscheme properties considered in this paper are {\em synonymies}
and {\em homonymies}. Given two x-components $x_A$ and $x_B$ belonging to different XML Schemas, a
{\em synonymy} between $x_A$ and $x_B$ indicates that they represent the same concept; an {\em
homonymy} between $x_A$ and $x_B$ denotes that they indicate different concepts yet having the same
name.

Our technique for computing interscheme properties is semantic
\cite{CaDeDe01,FaKrNe91,Ursino-TKDE1} in that, in order to determine the meaning of an x-component,
it examines the ``context'' which it has been defined in. It requires the presence of a thesaurus
storing lexical synonymies existing among the terms of a language. In particular, it exploits the
English language and WordNet\footnote{Actually, in the prototype implementing our technique,
WordNet is accessed by a suitable API.} \cite{Miller95}. The technique first extracts all
synonymies and, then, exploits them for deriving homonymies.

\subsection{Derivation of synonymies}
\label{Similarities}

As previously pointed out, in order to verify if two x-components $x_{1_j}$, belonging to an XML
Schema $S_1$, and $x_{2_k}$, belonging to an XML Schema $S_2$, are synonymous, it is necessary to
examine their neighborhoods. In particular, our approach operates as follows.

First it considers $neighborhood(x_{1_j},0)$ and $neighborhood(x_{2_k},0)$ and determines if they
are similar. This decision is made by computing the objective function associated with the maximum
weight matching of a suitable bipartite graph constructed from the x-components of
$neighborhood(x_{1_j},0)$ and $neighborhood(x_{2_k},0)$ and their lexical synonymies as stored in
the thesaurus (see below for all details). If $neighborhood(x_{1_j},0)$ and
$neighborhood(x_{2_k},0)$ are similar it is possible to conclude that $x_{1_j}$ and $x_{2_k}$ are
{\em synonymous} \cite{CaDeDe01,Ursino-TKDE1}. However, observe that $neighborhood(x_{1_j},0)$
(resp., $neighborhood(x_{2_k},0)$) takes into account only attributes and simple elements of
$x_{1_j}$ (resp., $x_{2_k}$); therefore, it considers quite a limited context. As a consequence,
the synonymy between $x_{1_j}$ and $x_{2_k}$ derived in this case is more ``syntactic'' than
``semantic'' \cite{FaKrNe91,CaDeDe01,Ursino-TKDE1}.

If we need a more ``severe'' level of synonymy detection it is necessary to require not only the
similarity of $neighborhood(x_{1_j},0)$ and $neighborhood(x_{2_k},0)$ but also that of the other
neighborhoods of $x_{1_j}$ and $x_{2_k}$. More specifically, it is possible to introduce a
``severity'' level $u$ at which synonymies are derived and to say that $x_{1_j}$ and $x_{2_k}$ are
synonymous with severity level equal to $u$ if $neighborhood(x_{1_j},v)$ is similar to
$neighborhood(x_{2_k},v)$ for each $v$ less than or equal to $u$. The following proposition states
an upper bound to the severity level that can be specified for x-component synonymy derivation.

\begin{proposition}
\label{maximum-level}

{\em Let $S_1$ and $S_2$ be two XML documents; let $x_{1_j}$ (resp., $x_{2_k}$) be an x-component of $S_1$
(resp., $S_2$); finally, let $m$ be the maximum between the number of complex elements of $S_1$ and $S_2$.
The maximum severity level possibly existing for the synonymy between $x_{1_j}$ and $x_{2_k}$ is $m-1$.}
$\punto$
\end{proposition}

\noindent A function $synonymous$ can be defined which receives two x-components $x_{1_j}$ and
$x_{2_k}$ and an integer $u$ and returns $true$ if $x_{1_j}$ and $x_{2_k}$ are synonymous with a
severity level equal to $u$, $false$ otherwise.

As previously pointed out, computing the synonymy between two x-compo-nents $x_{1_j}$ and $x_{2_k}$
implies determining when two neighborhoods are similar. In order to carry out such a task, it is
necessary to compute the objective function associated with the maximum weight matching relative to
a specific bipartite graph obtained from the x-components of the neighborhoods into consideration.

More specifically, let $BG(x_{1_j}, x_{2_k}, u) = \langle N(x_{1_j}, x_{2_k}, u), A(x_{1_j},
x_{2_k}, u) \rangle$ be the bipartite graph associated with $neighborhood(x_{1_j},u)$ and
$neighborhood(x_{2_k},u)$ (in the following we shall use the notation $BG(u)$ instead of
$BG(x_{1_j}, x_{2_k}, u)$ when this is not confusing). In $BG(u)$, $N(u) = P(u) \cup Q(u)$
represents the set of nodes; there is a node in $P(u)$ (resp., $Q(u)$) for each x-component of
$neighborhood(x_{1_j},u)$ (resp., $neighborhood(x_{2_k},u)$). $A(u)$ is the set of arcs; there is
an arc between $p_e \in P(u)$ and $q_f \in Q(u)$ if a synonymy between the names of the
x-components associated with $p_e$ and $q_f$ holds in the reference thesaurus. The maximum weight
matching for $BG(u)$ is a set $A'(u) \subseteq A(u)$ of edges such that, for each node $x \in P(u)
\cup Q(u)$, there is at most one edge of $A'(u)$ incident onto $x$ and $|A'(u)|$ is maximum (for
algorithms solving the maximum weight matching problem, see \cite{Galil86}). The objective function
we associate with the maximum weight matching is $\phi_{BG}(u) = \frac{2 |A'(u)|}{|P(u)| + |Q(u)|}
$.

We assume that if $\phi_{BG}(u) > \frac{1}{2}$ then $neighborhood(x_{1_j},u)$ and \\
$neighborhood(x_{2_k},u)$ are similar; otherwise they are dissimilar. Such an assumption derives
from the consideration that two sets of objects can be considered similar if the number of similar
components is greater than the number of the dissimilar ones or, in other words, if the number of
similar components is greater than half of the total number of components.

We present now the following theorem stating the computational complexity of the x-components'
similarity extraction.

\begin{theorem}
\label{Similar-Computation-Cost} {\em Let $S_1$ and $S_2$ be two XML documents. Let $x_{1_j}$ (resp.,
$x_{2_k}$) be an x-component of $S_1$ (resp., $S_2$). Let $u$ be the selected severity level. Finally, let
$p$ be the maximum between the cardinality of $neighborhood(x_{1_j},u)$ and $neighborhood(x_{2_k},u)$. The
worst case time complexity for computing $synony$-$mous(x_{1_j}, x_{2_k}, u)$ is $O((u+1) \times p^3)$.}
$\punto$
\end{theorem}

\begin{corollary}
{\em  Let $S_1$ and $S_2$ be two XML documents. Let $u$ be the severity level. Let $m$ be the
maximum between the number of complex elements of $S_1$ and $S_2$. Finally, let $q$ be the maximum
cardinality relative to a neighborhood of $S_1$ or $S_2$. The worst case time complexity for
deriving all synonymies existing, at the severity level $u$, between $S_1$ and $S_2$ is $O((u+1)
\times q^3 \times m^2)$}. $\punto$

\end{corollary}

\subsection{Derivation of homonymies}
\label{Type-Relationships}

After synonymies among x-components of $S_1$ and $S_2$ have been extracted, ho-monymies can be
directly derived from them. More specifically, we say that an homonymy holds between $x_{1_j}$ and
$x_{2_k}$ with a severity level equal to $u$ if $synonymous(x_{1_j}, x_{2_k}, u) = false$ and both
$x_{1_j}$ and $x_{2_k}$ have the same name.

It is possible to define a boolean function {\em homonymous}, which receives two x-components
$x_{1_j}$ and $x_{2_k}$ and an integer $u$ and returns $true$ if there exists an homonymy between
$x_{1_j}$ and $x_{2_k}$ with a severity level equal to $u$; {\em homonymous} returns $false$
otherwise.

\begin{example}
\label{interscheme-prop-example}

\noindent Consider the XML Schemas $S_1$ and $S_2$, shown in Figures \ref{S1} and \ref{S2}.
Consider also the x-components $customer_{[S_1]}$\footnote{Here and in the following, we use the
notation $x_{[S]}$ to indicate the x-component $x$ of the XML Schema $S$.} and $client_{[S_2]}$. In
order to check if they are synonymous with a severity level $0$, it is necessary to compute the
function $synonymous(customer_{[S_1]},client_{[S_2]},0)$. Now, $neighborhood(customer_{[S_1]},$
\\ $0)$ has been shown in Example \ref{neighborhood-example}; as for $neighborhood(client_{[S_2]},0)$,
we have:

\vspace*{0.2cm}

{\em neighborhood($client_{[S_2]}$,0) = \{$client_{[S_2]}$, $SSN_{[S_2]}$, $firstName_{[S_2]}$,

\hfill{$lastName_{[S_2]}$, $address_{[S_2]}$, $phone_{[S_2]}$, $email_{[S_2]}$\}}}

\hspace*{0.2cm}

\begin{figure*}[h]
{\tiny
\begin{minipage} [t] {6cm}
\begin{verbatim}

<?xml version="1.0" encoding="UTF-8"?>
<xs:schema xmlns:xs="http://www.w3.org/2001/XMLSchema">
    <!-- Definition of attributes -->
    <xs:attribute name="SSN" type="xs:string"/>
    <xs:attribute name="code" type="xs:ID"/>
    <xs:attribute name="purchasedCDDAs" type="xs:IDREFS"/>
    <xs:attribute name="purchasedMiniDisks" type="xs:IDREFS"/>
    <xs:attribute name="purchaseDate" type="xs:date"/>
    <xs:attribute name="quantity" type="xs:integer"/>
    <xs:attribute name="bitRate" type="xs:integer"/>
    <!-- Definition of simple elements -->
    <xs:element name="firstName" type="xs:string"/>
    <xs:element name="lastName" type="xs:string"/>
    <xs:element name="address" type="xs:string"/>
    <xs:element name="phone" type="xs:string"/>
    <xs:element name="email" type="xs:string"/>
    <xs:element name="artist" type="xs:string"/>
    <xs:element name="title" type="xs:string"/>
    <xs:element name="song" type="xs:string"/>
    <xs:element name="year" type="xs:integer"/>
    <xs:element name="genre" type="xs:string"/>
    <!-- Definition of complex elements -->
    <xs:element name="CDDAPurchase">
        <xs:complexType>
            <xs:attribute ref="purchaseDate"/>
            <xs:attribute ref="purchasedCDDAs"/>
        </xs:complexType>
    </xs:element>
    <xs:element name="miniDiskPurchase">
        <xs:complexType>
            <xs:attribute ref="purchaseDate"/>
            <xs:attribute ref="purchasedMiniDisks"/>
        </xs:complexType>
    </xs:element>
    <xs:element name="client">
        <xs:complexType>
            <xs:sequence>
                <xs:element ref="firstName"/>
                <xs:element ref="lastName"/>
                <xs:element ref="address"/>
                <xs:element ref="phone" minOccurs="0"
                    maxOccurs="unbounded"/>
                <xs:element ref="email" minOccurs="0"
                    maxOccurs="unbounded"/>
                <xs:element ref="CDDAPurchase" minOccurs="0"
                    maxOccurs="unbounded"/>
\end{verbatim}
\end{minipage}$ $
\begin{minipage}[t]{6cm}
\begin{verbatim}

                <xs:element ref="miniDiskPurchase" minOccurs="0"
                    maxOccurs="unbounded"/>
            </xs:sequence>
            <xs:attribute ref="SSN" use="required"/>
        </xs:complexType>
    </xs:element>
    <xs:element name="CDDA">
        <xs:complexType>
            <xs:attribute ref="code" use="required"/>
            <xs:attribute ref="quantity"/>
        </xs:complexType>
    </xs:element>
    <xs:element name="miniDisk">
        <xs:complexType>
            <xs:attribute ref="code" use="required"/>
            <xs:attribute ref="quantity"/>
            <xs:attribute ref="bitRate"/>
        </xs:complexType>
    </xs:element>
    <xs:element name="composition">
        <xs:complexType>
            <xs:sequence>
                <xs:element ref="artist" maxOccurs="unbounded"/>
                <xs:element ref="title"/>
                <xs:element ref="song" maxOccurs="unbounded"/>
                <xs:element ref="year"/>
                <xs:element ref="genre"/>
                <xs:element ref="CDDA" minOccurs="0"/>
                <xs:element ref="miniDisk" minOccurs="0"/>
            </xs:sequence>
        </xs:complexType>
    </xs:element>
    <!-- Definition of root element -->
    <xs:element name="store">
        <xs:complexType>
            <xs:sequence>
                <xs:element ref="client" maxOccurs="unbounded"/>
                <xs:element ref="composition" maxOccurs="unbounded"/>
            </xs:sequence>
        </xs:complexType>
    </xs:element>
</xs:schema>

\end{verbatim}
\end{minipage}
} \caption{The XML Schema $S_2$} \label{S2}
\end{figure*}

\noindent The function $\phi_{BG}(0)$ computed by $synonymous$ in this case is  $\frac{2
|A'(0)|}{|P(0)| + |Q(0)|} = \frac{2 \times 5}{8+7}=0.67 > \frac{1}{2}$; therefore
$synonymous(customer_{[S_1]},client_{[S_2]},0)=true$.

In an analogous way, $synonymous(customer_{[S_1]}, client_{[S_2]}, 1)$ can be computed. In
particular, in this case, $\phi_{BG}(1) = 0.43 < \frac{1}{2}$; as a consequence, \\
$synonymous(customer_{[S_1]},client_{[S_2]},1)=false$, i.e. $customer_{[S_1]}$ and $client_{[S_2]}$
cannot be considered synonymous with a severity level 1.

All the other synonymies can be derived analogously. As for these Schemas, no homonymy has been
found. $\punto$
\end{example}

\section{The Integration Task}
\label{Integration-Task}

In this section we propose an integration algorithm which receives two XML Schemas $S_1$ and $S_2$
and a severity level $u$ and returns the integrated XML Schema $S_G$. The algorithm consists of two
steps, namely: {\em (i)} construction of a {\em Merge Dictionary} $MD(u)$ and a {\em Rename
Dictionary} $RD(u)$; {\em (ii)} exploitation of $MD(u)$ and $RD(u)$ for obtaining the global
Schema.

Preliminarily it is necessary to observe that in XML Schemas there exists a large variety of data
types. Some of them, e.g. {\em Byte} and {\em Int}, are compatible in the sense that each attribute
or simple element whose type is {\em Byte} can be treated as an attribute or a simple element whose
type is {\em Int}; in this case {\em Int} is said {\em more general} than {\em Byte}. Other types,
e.g. {\em Int} and {\em Date}, are not compatible. Compatibility rules are analogous to the
corresponding ones valid for high level programming languages.

\subsection{Construction of MD(u) and RD(u)}
\label{MD-RD}

At the end of interscheme property derivation, it could happen that an x-component of a Schema is
synonymous (resp., homonymous) with more than one x-components of the other Schema. The integration
algorithm we are proposing here needs each x-component of a Schema to be synonymous (resp.,
homonymous) with at most one x-component of the other Schema. In order to satisfy this requirement,
it is necessary to construct a {\em Merge Dictionary} $MD(u)$ and an {\em Rename Dictionary}
$RD(u)$ by suitably filtering previously derived synonymies and homonymies.

The construction of $MD(u)$ begins with the definition of a support bipartite graph $SimG(u) =
\langle SimNSet_1(u) \cup SimNSet_2(u), SimASet(u) \rangle$.

There is a node $n_{1_j}$ (resp., $n_{2_k}$) in $SimNSet_1(u)$ (resp., $SimNSet_2(u)$) for each
complex element $E_{1_j}$ (resp., $E_{2_k}$) belonging to $S_1$ (resp., $S_2$). There is an arc
$A_{jk} = \langle n_{1_j}, n_{2_k} \rangle \in SimASet(u)$ if $synonymous(E_{1_j}, E_{2_k},
u)=true$; the label of each arc $A_{jk}$ is $f(n_{1_j}, n_{2_k})$ where:

\begin{quote}

\vspace*{-0.3cm}

$f(n_{1_j},n_{2_k}) = \left\{
\begin{array}{ll}
\phi_{BG}(E_{1_j},E_{2_k},u) & \hspace*{0.3cm} \mbox{if $A_{jk} \in SimASet(u)$} \\
0 & \hspace*{0.3cm} \mbox{\em otherwise}
\end{array}
\right. $

\vspace*{-0.3cm}

\end{quote}

\noindent Function $f$ has been defined in such a way to maximize the sum of the similarity degrees
involving complex elements of $S_1$ and $S_2$.

After this, a maximum weight matching is computed on $SimG(u)$; this selects a subset
$SimASubSet(u) \subseteq SimASet(u)$ which maximizes the objective function $\phi_{Sim}(u) =
\sum_{\langle n_{1_j}, n_{2_k} \rangle \in SimASubSet(u)} f(n_{1_j},n_{2_k})$.

For each arc $A'_{jk} = \langle n'_{1_j}, n'_{2_k} \rangle \in SimASubSet(u)$ a pair $\langle
E'_{1_j}, E'_{2_k} \rangle$ is added to $MD(u)$.

In addition, let $E'_{1_j}$ (resp., $E'_{2_k}$) be a complex element of $S_1$ (resp., $S_2$) such
that $\langle E'_{1_j}, E'_{2_k} \rangle \in MD(u)$ and let $x'_{1_j}$ (resp., $x'_{2_k}$) be an
attribute or a simple element of $E'_{1_j}$ (resp., $E'_{2_k}$); then $\langle x'_{1_j}, x'_{2_k}
\rangle$ is added to $MD(u)$ if {\em (i)} a synonymy between the name of $x'_{1_j}$ and that of
$x'_{2_k}$ holds in the reference thesaurus and the data types of $x'_{1_j}$ and $x'_{2_k}$ are
compatible, or {\em (ii)} $x'_{1_j}$ and $x'_{2_k}$ have the same name, the same typology and
compatible data types.

After $MD(u)$ has been constructed, it is possible to derive $RD(u)$. More specifically, a pair of
x-components $\langle x''_{1_j}, x''_{2_k} \rangle$ is added to $RD(u)$ if $x''_{1_j}$ and
$x''_{2_k}$ are two elements or two attributes having the same name and $\langle x''_{1_j},
x''_{2_k} \rangle \not \in MD(u)$.

\subsection{Construction of the global XML Schema}
\label{XML-Schema-Construction}

After $MD(u)$ and $RD(u)$ have been derived, it is possible to exploit them for constructing a
global Schema $S_G$. Our integration algorithm assumes that $S_1$ and $S_2$ are represented in the
{\em referenced style}, i.e., that they consist of sequences of elements and that each element may
refer to other elements by means of the {\em ref} attribute. Actually, an XML Schema could be
defined in various other ways (e.g., with the {\em inline style}); however, simple rules can be
easily defined for translating it in the {\em referenced style} (see \cite{Thompson*01} for more
details on the various definition styles).

More formally, $S_1$ and $S_2$ can be represented as:

\begin{center}

\vspace*{-0.3cm}

$S_1 = \langle x_{1_1}, x_{1_2}, \ldots, x_{1_i}, \ldots, x_{1_n} \rangle$; $S_2 = \langle x_{2_1},
x_{2_2}, \ldots, x_{2_j}, \ldots, x_{2_m} \rangle$

\vspace*{-0.3cm}

\end{center}

\noindent where $x_{1_1}, \ldots, x_{1_n}, x_{2_1}, \ldots, x_{2_m}$ are x-components. A first,
rough, version of $S_G$ can be obtained by constructing a list containing all the x-components of
$S_1$ and $S_2$:

\vspace*{-0.3cm}

\begin{center}
$S_G = \langle x_{1_1}, \ldots, x_{1_n}, x_{2_1}, \ldots, x_{2_m} \rangle$
\end{center}

\vspace*{-0.3cm}

This version of $S_G$ could present some redundancies and/or ambiguities. In order to remove them
and, consequently, to refine $S_G$, $MD(u)$ and $RD(u)$ must be examined and some tasks must be
performed for each of the properties they store. More specifically, consider $MD(u)$ and let
$\langle E_{1_j}, E_{2_k} \rangle \in MD(u)$ be a synonymy between two complex elements. $E_{1_j}$
and $E_{2_k}$ are merged into a complex element $E_{jk}$. The name of $E_{jk}$ is one between the
names of $E_{1_j}$ and $E_{2_k}$. The set of sub-elements of $E_{jk}$ is obtained by applying the
{\tt xs:sequence} indicator to the sets of sub-elements of $E_{1_j}$ and $E_{2_k}$; the list of
attributes of $E_{jk}$ is formed by the attributes of $E_{1_j}$ and $E_{2_k}$. Note that, after
these tasks have been carried out, it could happen that:

\vspace*{-0.3cm}

\begin{itemize}

\item A tuple $\langle A_{jk}', A_{jk}'' \rangle$, such that $A_{jk}'$ and $A_{jk}''$ are
attributes of $E_{jk}$, belongs to $MD(u)$. In this case $A_{jk}'$ and $A_{jk}''$ are merged into
an attribute $A_{jk}^*$; the name of $A_{jk}^*$ is one between the names of $A_{jk}'$ and
$A_{jk}''$; the type of $A_{jk}^*$ is the most general one between those of $A_{jk}'$ and
$A_{jk}''$.

\item A tuple $\langle E_{jk}', E_{jk}'' \rangle$, such that $E_{jk}'$ and $E_{jk}''$ are simple
elements of $E_{jk}$, belongs to $MD(u)$. In this case $E_{jk}'$ and $E_{jk}''$ are merged into an
element $E_{jk}^*$; the name of $E_{jk}^*$ is one between the names of $E_{jk}'$ and $E_{jk}''$;
the type of $E_{jk}^*$ is the most general one between those of $E_{jk}'$ and $E_{jk}''$; the
$minOccurs$ (resp., the $maxOccurs$) indicator of $E_{jk}^*$ is the minimum (resp., the maximum)
between the corresponding ones relative to $E_{jk}'$ and $E_{jk}''$.

\item A tuple $\langle E_{jk}'', A_{jk}'' \rangle$, such that $E_{jk}''$ is a simple sub-element of
$E_{jk}$ and $A_{jk}''$ is an attribute of $E_{jk}$, belongs to $MD(u)$. In this case, $A_{jk}''$
is removed since its information content is equivalent to that of $E_{jk}''$ and the representation
of an information content by means of an element is more general than that obtained by exploiting
an attribute.

\end{itemize}

\vspace*{-0.3cm}

After this, all references to $E_{1_j}$ and $E_{2_k}$ in $S_G$ are transformed into references to
$E_{jk}$; the $maxOccurs$ and the $minOccurs$ indicators associated with $E_{jk}$ are derived from
the corresponding ones relative to $E_{1_j}$ and $E_{2_k}$ and, finally, $E_{1_j}$ is replaced by
$E_{jk}$ whereas $E_{2_k}$ is removed from $S_G$.

After $MD(u)$ has been examined, it is necessary to consider $RD(u)$; in particular, let $\langle
x_{1_j}, x_{2_k} \rangle$ be a tuple of $RD(u)$ such that $x_{1_j}$ and $x_{2_k}$ are both elements
or both attributes of the same element. In this case it is necessary to modify the name of either
$x_{1_j}$ or $x_{2_k}$ and all the corresponding references.

Observe that, after all these activities have been performed, $S_G$ could contain two root
elements. Such a situation occurs when the root elements $E_{1_r}$ of $S_1$ and $E_{2_r}$ of $S_2$
are not synonymous. In this case it is necessary to create a new root element $E_{G_r}$ in $S_G$
whose set of sub-elements is obtained by applying the {\tt xs:all} indicator to $E_{1_r}$ and
$E_{2_r}$. The occurrence indicators associated with $E_{1_r}$ and $E_{2_r}$ are $minOccurs=0$ and
$maxOccurs=1$.

As for the computational complexity of the integration task, it is possible to state the following
theorem.

\begin{theorem}
{\em Let $S_1$ and $S_2$ be two XML Schemas, let $n$ be the maximum between $|XCompSet(S_1)|$ and
$|XCompSet(S_2)|$ and let $m$ be the maximum between the number of complex elements of $S_1$ and the number
of complex elements of $S_2$. The worst case time complexity for integrating $S_1$ and $S_2$ into a global
Schema $S_G$ is $O(m \times n^2)$.} $\punto$

\end{theorem}

\begin{example}
\label{integration-example}

Assume a user wants to integrate the XML Schemas $S_1$ and $S_2$, shown in Figures \ref{S1} and
\ref{S2}, and the severity level she/he specifies is 0. $MD(0)$ is illustrated in Table \ref{MD0};
$RD(0)$ is empty because no homonymy has been found among x-components of $S_1$ and $S_2$ (see
Example \ref{interscheme-prop-example}). Initially a rough version of $S_G$ is constructed that
contains all the x-components of $S_1$ and $S_2$; the refined version of $S_G$ is obtained by
removing (possible) redundancies and/or ambiguities present therein.

\begin{table}[t]
\begin{center}
{\scriptsize
\begin{tabular}{||c|c||c||c|c||}
\hline \hline
{\em x-component of $S_1$} & {\em x-component of $S_2$} & & {\em x-component of $S_1$} & {\em x-component of $S_2$} \\
\hline \hline
shop & store & \hspace*{0.5cm} & customer & client \\
music & composition & & SSN & SSN \\
firstName & firstName & & lastName & lastName \\
address & address & & code & code \\
artist & artist & & title & title \\
pubYear & year & & genre & genre\\
\hline \hline
\end{tabular}
}
\end{center}
\caption{The Merge Dictionary $MD(0)$} \label{MD0}
\end{table}

The first step of the refinement phase examines all synonymies among complex elements stored in
$MD(0)$. As an example, consider the synonymous elements {\em customer$_{[S_1]}$} and {\em
client}$_{[S_2]}$; they must be merged in one single element. This task is carried out as follows.
First a new element {\em customer}$_{[S_G]}$ is created in $S_G$. The set of sub-elements of {\em
customer}$_{[S_G]}$ is obtained by applying the {\tt xs:sequence} indicator to the sets of
sub-elements of $customer_{[S_1]}$ and $client_{[S_2]}$; the list of attributes of {\em
customer}$_{[S_G]}$ is formed by the attributes of {\em customer}$_{[S_1]}$ and {\em
client}$_{[S_2]}$. At the end of this task, {\em customer}$_{[S_G]}$ contains two attributes named
{\em SSN}. Since the tuple $\langle SSN, SSN \rangle$ belongs to $MD(0)$, the two attributes are
merged into a single attribute {\em SSN} having type ``string''. An analogous procedure is applied
to sub-element pairs $\langle firstName_{[S_1]}, firstName_{[S_2]} \rangle$, $\langle${\em
lastName}$_{[S_1]}$, {\em lastName}$_{[S_2]} \rangle$ and $\langle${\em address}$_{[S_1]}$, {\em
address}$_{[S_2]}\rangle$.

After this, all references to {\em customer}$_{[S_1]}$ and {\em client}$_{[S_2]}$ are transformed
into references to {\em customer}$_{[S_G]}$; finally, {\em customer}$_{[S_1]}$ is replaced by {\em
customer}$_{[S_G]}$ where-as {\em client$_{[S_2]}$} is removed from $S_G$. All the other synonymies
stored in $MD(0)$ are handled similarly. Since no homonymy has been found, no further action is
necessary. The global XML Schema $S_G$, obtained at the end of the integration activity, is shown
in Figure \ref{SG}.

\begin{figure*}[p]
{\tiny
\begin{minipage}[t]{6cm}
\begin{verbatim}

<?xml version="1.0" encoding="UTF-8"?>
<xs:schema xmlns:xs="http://www.w3.org/2001/XMLSchema">
    <!-- Definition of attributes -->
    <xs:attribute name="SSN" type="xs:string"/>
    <xs:attribute name="code" type="xs:ID"/>
    <xs:attribute name="acquiredBooks" type="xs:IDREFS"/>
    <xs:attribute name="acquiredMusics" type="xs:IDREFS"/>
    <xs:attribute name="acquirementDate" type="xs:date"/>
    <xs:attribute name="purchasedCDDAs" type="xs:IDREFS"/>
    <xs:attribute name="purchasedMiniDisks" type="xs:IDREFS"/>
    <xs:attribute name="purchaseDate" type="xs:date"/>
    <xs:attribute name="quantity" type="xs:integer"/>
    <xs:attribute name="bitRate" type="xs:integer"/>
    <!-- Definition of simple elements -->
    <xs:element name="firstName" type="xs:string"/>
    <xs:element name="lastName" type="xs:string"/>
    <xs:element name="address" type="xs:string"/>
    <xs:element name="gender" type="xs:string"/>
    <xs:element name="birthDate" type="xs:date"/>
    <xs:element name="profession" type="xs:string"/>
    <xs:element name="phone" type="xs:string"/>
    <xs:element name="email" type="xs:string"/>
    <xs:element name="artist" type="xs:string"/>
    <xs:element name="author" type="xs:string"/>
    <xs:element name="title" type="xs:string"/>
    <xs:element name="song" type="xs:string"/>
    <xs:element name="pubYear" type="xs:integer"/>
    <xs:element name="publisher" type="xs:string"/>
    <xs:element name="genre" type="xs:string"/>
    <xs:element name="support" type="xs:string"/>
    <!-- Definition of complex elements -->
    <xs:element name="bookAcquirement">
        <xs:complexType>
            <xs:attribute ref="acquirementDate"/>
            <xs:attribute ref="acquiredBooks"/>
        </xs:complexType>
    </xs:element>
    <xs:element name="musicAcquirement">
        <xs:complexType>
            <xs:attribute ref="acquirementDate"/>
            <xs:attribute ref="acquiredMusics"/>
        </xs:complexType>
    </xs:element>
    <xs:element name="book">
        <xs:complexType>
            <xs:sequence>
                <xs:element ref="author" maxOccurs="unbounded"/>
                <xs:element ref="title"/>
                <xs:element ref="publisher"/>
                <xs:element ref="pubYear"/>
                <xs:element ref="genre"/>
            </xs:sequence>
            <xs:attribute ref="code" use="required"/>
        </xs:complexType>
    </xs:element>
    <xs:element name="CDDAPurchase">
        <xs:complexType>
            <xs:attribute ref="purchaseDate"/>
            <xs:attribute ref="purchasedCDDAs"/>
        </xs:complexType>
    </xs:element>
    <xs:element name="miniDiskPurchase">
        <xs:complexType>
            <xs:attribute ref="purchaseDate"/>
            <xs:attribute ref="purchasedMiniDisks"/>
        </xs:complexType>
    </xs:element>

\end{verbatim}
\end{minipage}$ $
\begin{minipage}[t]{6cm}
\begin{verbatim}

    <xs:element name="customer">
        <xs:complexType>
            <xs:sequence>
                <xs:element ref="firstName"/>
                <xs:element ref="lastName"/>
                <xs:element ref="address"/>
                <xs:element ref="gender"/>
                <xs:element ref="birthDate"/>
                <xs:element ref="profession"/>
                <xs:element ref="bookAcquirement"
                    minOccurs="0" maxOccurs="unbounded"/>
                <xs:element ref="musicAcquirement"
                    minOccurs="0" maxOccurs="unbounded"/>
                <xs:element ref="phone"
                    minOccurs="0" maxOccurs="unbounded"/>
                <xs:element ref="email"
                    minOccurs="0" maxOccurs="unbounded"/>
                <xs:element ref="CDDAPurchase"
                    minOccurs="0" maxOccurs="unbounded"/>
                <xs:element ref="miniDiskPurchase"
                    minOccurs="0" maxOccurs="unbounded"/>
            </xs:sequence>
            <xs:attribute ref="SSN" use="required"/>
        </xs:complexType>
    </xs:element>
    <xs:element name="CDDA">
        <xs:complexType>
            <xs:attribute ref="code"/>
            <xs:attribute ref="quantity"/>
        </xs:complexType>
    </xs:element>
    <xs:element name="miniDisk">
        <xs:complexType>
            <xs:attribute ref="code"/>
            <xs:attribute ref="quantity"/>
            <xs:attribute ref="bitRate"/>
        </xs:complexType>
    </xs:element>
    <xs:element name="music">
        <xs:complexType>
            <xs:sequence>
                <xs:element ref="artist" maxOccurs="unbounded"/>
                <xs:element ref="title"/>
                <xs:element ref="pubYear"/>
                <xs:element ref="genre"/>
                <xs:element ref="support" minOccurs="0"/>
                <xs:element ref="song"
                    minOccurs="0" maxOccurs="unbounded"/>
                <xs:element ref="CDDA" minOccurs="0"/>
                <xs:element ref="miniDisk" minOccurs="0"/>
            </xs:sequence>
            <xs:attribute ref="code"/>
        </xs:complexType>
    </xs:element>
    <!-- Definition of root element -->
    <xs:element name="shop">
        <xs:complexType>
            <xs:sequence>
                <xs:element ref="customer"
                    maxOccurs="unbounded"/>
                <xs:element ref="music"
                    maxOccurs="unbounded"/>
                <xs:element ref="book"
                    minOccurs="0" maxOccurs="unbounded"/>
            </xs:sequence>
        </xs:complexType>
    </xs:element>
</xs:schema>

\end{verbatim}
\end{minipage}
} \caption{The integrated XML Schema $S_G$} \label{SG}
\end{figure*}

$\punto$
\end{example}

\section{Experiments}
\label{Experiments}

To test the performances of our approach we have carried out various experiments; these have been
performed on several XML Schemas taken from different application contexts. Involved XML Schemas
were very heterogeneous in their dimensions; indeed, the number of x-components associated with
them ranged from tens to hundreds.

The first series of experiments has been conceived for measuring correctness and completeness of
our interscheme property derivation algorithm. In particular, {\em correctness} lists the
percentage of properties returned by our techniques agreeing with those provided by humans; {\em
completeness} lists the percentage of properties returned by our approach with regard to the set of
properties provided by humans.

In more detail, we proceeded as follows: {\em (i)} we ran our algorithms on several pairs of XML
Schemas and collected the returned results; {\em (ii)} for each pair of Schemas we asked humans to
specify a set of significant interscheme properties; {\em (iii)} we computed the overall quality
figures by comparing the set of properties obtained as described at points 1 and 2 above.

As for severity level 0, we have obtained a correctness equal to 0.88 and a completeness equal to
1,00.

Actually, the intrinsic characteristics of our algorithm led us to think that, if the severity
level increases, the correctness increases as well, whereas the completeness decreases. In order to
verify this idea, we have performed a second series of experiments devoted to measure correctness
and completeness in presence of variations of the severity level. Table
\ref{Correctness-Completeness} shows obtained results up to a severity level equal to 3; for higher
severity levels, variations of correctness and completeness are not significant.

\begin{table}
\begin{center}
{\scriptsize
\begin{tabular}{||c|c|c||}
\hline \hline
{\em Severity Level} & {\em Correctness} & {\em Completeness} \\
\hline \hline
Level 0 & 0.88 & 1.00 \\
Level 1 & 0.97 & 0.81 \\
Level 2 & 0.97 & 0.78 \\
Level 3 & 0.97 & 0.73 \\
\hline \hline
\end{tabular}
}
\end{center}
\caption{Correctness and Completeness of our approach at various severity levels}
\label{Correctness-Completeness}
\end{table}

Results presented in Table \ref{Correctness-Completeness} confirmed our intuitions. Indeed, at
severity level 1, correctness increases of a factor of 9\% whereas completeness decreases of a
factor of 19\% w.r.t. correctness and completeness relative to severity level 0. As for severity
levels greater than 1, we have verified that correctness does not increase whereas completeness
slightly decreases w.r.t. level 1.

In our opinion such a result is extremely relevant; indeed, it allows us to conclude that, in
informal situations, the right severity level is 0 whereas, in more formal contexts, the severity
level must be at least 1.

After this, we have computed variations of the time required for deriving interscheme properties
caused by an increase of the severity level. Obtained results are shown in Table \ref{Time}. In the
table the value associated with severity level $i$ ($1 \leq i \leq 3$) is to be intended as the
percentage of time additionally required w.r.t. severity level $i-1$.

\begin{table}
\begin{center}
{\scriptsize
\begin{tabular}{||c|c||}
\hline \hline
{\em Severity Level} & {\em Time Increase} \\
\hline \hline
Level 1 & 56\% \\
Level 2 & 14\% \\
Level 3 & 20\% \\
\hline \hline
\end{tabular}
}
\end{center}
\caption{Increase of the time required by our approach at various severity levels} \label{Time}
\end{table}

Table \ref{Time} shows that the increase of time required for computing interscheme properties when
the algorithm passes from the severity level 0 to the severity level 1 is significant. Vice versa,
further severity level increases do not lead to significant increases of the time necessary for
computing interscheme properties. This observation further confirms results obtained by the
previous experiments, i.e., that the most relevant differences in the results obtained by applying
our approach can be found between the severity levels 0 and 1.

\section{Related Work}
\label{Related-Work}

In the literature many approaches for performing interscheme property extraction and data source
integration have been proposed. Even if they are quite numerous and various, to the best of our
knowledge, none of them guarantees the possibility to choose a ``severity'' level against which the
various activities are carried out. In this section we examine some of these approaches and
highlight their similarities and differences w.r.t. our own.

In \cite{Passi*02} an XML Schema integration framework is proposed. It consists of three phases, namely
pre-integration, comparison and integration. After this, {\em conflict resolution} and {\em restructuring}
are performed for obtaining the global refined Schema. To the best of our knowledge the approach of
\cite{Passi*02} is the closest to our own. In particular, {\em (i)} both of them are {\em rule-based}
\cite{RaBe01}; {\em (ii)} both of them assume that the global Schema is formulated in a {\em referenced
style} rather than in an {\em inline style} (see \cite{Thompson*01} for more details); {\em (iii)}
integration rules proposed in \cite{Passi*02} are quite similar to those characterizing our approach. The
main differences existing between them are the following: {\em (i)} the approach of \cite{Passi*02} requires
a preliminary translation of an XML Schema into an $XSDM$ Schema; such a translation is not required by our
approach; {\em (ii)} the integration task in \cite{Passi*02} is graph-based and object-oriented whereas, in
our approach, it is directly based on {\em x-components};.

In \cite{Lee*02} the system {\em XClust} is presented whose purpose is XML data source integration. More
specifically, {\em XClust} determines the similarity degrees of a group of DTD's by considering not only the
corresponding linguistic and structural information but also their semantics. It is possible to recognize
some similarities between our approach and {\em XClust}; in particular, {\em (i)} both of them have been
specifically conceived for operating on XML data sources (even if our approach manages XML Schemas whereas
{\em XClust} operates on DTD's); {\em (ii)} both of them consider not only linguistic similarities but also
semantic ones. There are also several differences between the two approaches; more specifically, {\em (i)} to
perform the integration activity, {\em XClust} requires the support of a hierarchical clustering whereas our
approach adopts schema matching techniques; {\em (ii)} {\em XClust} represents DTD's as trees; as a
consequence, element neighborhoods are quite different from those constructed by our approach; {\em (iii)}
{\em XClust} exploits some weights and thresholds whereas our approach does not use them; as a consequence,
{\em XClust} provides more refined results but these last are strongly dependent on the correctness of a
tuning phase devoted to set weights and thresholds.

In \cite{MeRaBe03} the system {\em Rondo} is presented. It has been conceived for integrating and
manipulating relational schemas, XML Schemas and SQL views. {\em Rondo} exploits a graph-based approach for
modeling information sources and a set of high-level operators for matching obtained graphs. Rondo uses the
{\em Similarity Flooding Algorithm}, a graph-matching algorithm proposed in \cite{MeGaRa02}, to perform
schema matching activity. Finally, it merges involved information sources according to three steps: Node
Renaming, Graph Union and Conflict Resolution. There are important similarities between {\em Rondo} and our
approach; indeed both of them are semi-automatic and exploit schema matching techniques. The main differences
existing between them are the following: {\em (i)} {\em Rondo} is generic, i.e., it can handle various kinds
of information sources; vice versa our approach is specialized for XML Schemas; {\em (ii)} {\em Rondo} models
involved information sources as graphs whereas our approach directly operates on XML Schemas; {\em (iii)}
{\em Rondo} exploits a sophisticated technique (i.e., the Similarity Flooding Algorithm) for carrying out
schema matching activities \cite{MeGaRa02}; as a consequence, it obtains very precise results but is
time-expensive and requires a heavy human feedback; on the contrary, our approach is less sophisticated but
is well suited when involved information sources are numerous and large.

In \cite{SaCaHe02} an XML-based integration approach, capable of handling various source formats,
is presented. Both this approach and our own operate on XML documents and carry out a semantic
integration. However, {\em (i)} the approach of \cite{SaCaHe02} operates on DTD's and requires to
translate them in an appropriate formalism called ORM/NIAM \cite{Halpin98}; vice versa, our
approach directly operates on XML Schemas; {\em (ii)} the global Schema constructed by the approach
of \cite{SaCaHe02} is represented in the ORM/NIAM formalism whereas our approach direcly returns a
global XML Schema; {\em (iii)} the approach of \cite{SaCaHe02} is quite complex to be applied when
involved sources are numerous.

In \cite{RoMy01} the DIXSE (Data Integration for XML based on Schematic Knowledge) tool is
presented, aiming at supporting the integration of a set of XML documents. Both DIXSE and our
approach are semantic and operate on XML documents; both of them exploit structural and
terminological relationships for carrying out the integration activity. The main differences
between them reside in the interscheme property extraction technique; indeed, DIXSE requires the
support of the user whereas our approach derives them almost automatically. As a consequence,
results returned by DIXSE could be more precise than those provided by our approach but, when the
number of sources to integrate is high, the effort DIXSE requires to the user might be particularly
heavy.

In \cite{DoDoHa01} a {\em machine learning} approach, named {\sffamily LSD} (Learning Source
Description), for carrying out schema matching activities, is proposed. It has been extended also
to ontologies in {\sffamily GLUE} \cite{Doan*02}. {\sffamily LSD} requires quite a heavy support of
the user during the initial phase, for carrying out training tasks; however, after this phase, no
human intervention is required. Both {\sffamily LSD} and our approach operate mainly on XML
sources. They differ especially in their purposes; indeed, {\sffamily LSD} aims at deriving
interscheme properties whereas our approach has been conceived mainly for handling integration
activities. In addition, as far as interscheme property derivation is concerned, it is worth
observing that {\sffamily LSD} is {\em ``learner-based''} whereas our approach is {\em
``rule-based''} \cite{RaBe01}. Finally, {\sffamily LSD} requires a heavy human intervention at the
beginning and, then, is automatic; vice versa, our approach does not need a pre-processing phase
but requires the human intervention at the end for validating obtained results.

In \cite{DoRa02} the authors propose COMA (COmbining MAtch), an interactive and iterative system
for combining various schema matching approaches. The approach of COMA appears orthogonal to our
own; in particular, our approach could inherit some features from COMA (as an example, the idea of
operating iteratively) for improving the accuracy of its results. As for an important difference
between the two approaches, we observe that COMA is generic, since it handles a large variety of
information source formats; vice versa, our approach has been specifically conceived to handle XML
documents. In addition, our approach requires the user to specify only the {\em severity level};
vice versa, in COMA, the user must specify the {\em matching strategy} (i.e., the desired matchers
to exploit and the modalities for combining their results).

\section{Conclusions}
\label{Conclusions}

In this paper we have proposed an approach for the integration of a set of XML Schemas. We have
shown that our approach is specialized for XML documents, is almost automatic, semantic and
``light'' and allows the choice of the ``severity'' level against which the integration activity
must be performed. We have also illustrated some experiments we have carried out to test its
computational performances and the quality of results it obtains. Finally, we have examined various
other related approaches previously proposed in the literature and we have compared them with ours
by pointing out similarities and differences.

In the future we plan to exploit our approach in various other contexts typically benefiting of
information source integration, such as Cooperative Information Systems, Data Warehousing, Semantic
Query Processing and so on.

\end{document}

%% file: new_fig.tex
\def\PsfigVersion{1.9}
\ifx\undefined\psfig\else \fi

\let\LaTeXAtSign=\@
\let\@=\relax
\edef\psfigRestoreAt{\catcode`\@=\number\catcode`@\relax}
\catcode`\@=11\relax
\newwrite\@unused
\def\ps@typeout#1{{\let\protect\string\immediate\write\@unused{#1}}}
\ps@typeout{psfig/tex \PsfigVersion}

\def\figurepath{./}

\def\@nnil{\@nil}
\def\@empty{}
\def\@psdonoop#1\@@#2#3{}
\def\@psdo#1:=#2\do#3{\edef\@psdotmp{#2}\ifx\@psdotmp\@empty \else
    \expandafter\@psdoloop#2,\@nil,\@nil\@@#1{#3}\fi}
\def\@psdoloop#1,#2,#3\@@#4#5{\def#4{#1}\ifx #4\@nnil \else
       #5\def#4{#2}\ifx #4\@nnil \else#5\@ipsdoloop #3\@@#4{#5}\fi\fi}
\def\@ipsdoloop#1,#2\@@#3#4{\def#3{#1}\ifx #3\@nnil 
       \let\@nextwhile=\@psdonoop \else
      #4\relax\let\@nextwhile=\@ipsdoloop\fi\@nextwhile#2\@@#3{#4}}
\def\@tpsdo#1:=#2\do#3{\xdef\@psdotmp{#2}\ifx\@psdotmp\@empty \else
    \@tpsdoloop#2\@nil\@nil\@@#1{#3}\fi}
\def\@tpsdoloop#1#2\@@#3#4{\def#3{#1}\ifx #3\@nnil 
       \let\@nextwhile=\@psdonoop \else
      #4\relax\let\@nextwhile=\@tpsdoloop\fi\@nextwhile#2\@@#3{#4}}
\ifx\undefined\fbox
\newdimen\fboxrule
\newdimen\fboxsep
\newdimen\ps@tempdima
\newbox\ps@tempboxa
\long\def\fbox#1{\leavevmode\setbox\ps@tempboxa\hbox{#1}\ps@tempdima\fboxrule
    \advance\ps@tempdima \fboxsep \advance\ps@tempdima \dp\ps@tempboxa
   \hbox{\lower \ps@tempdima\hbox
  {\vbox{\hrule height \fboxrule
          \hbox{\vrule width \fboxrule \hskip\fboxsep
          \vbox{\vskip\fboxsep \box\ps@tempboxa\vskip\fboxsep}\hskip 
                 \fboxsep\vrule width \fboxrule}
                 \hrule height \fboxrule}}}}
\fi
\newread\ps@stream
\newif\ifnot@eof       
\newif\if@noisy        
\newif\if@atend        
\newif\if@psfile       
{\catcode`\%=12\global\gdef\epsf@start{
\def\epsf@PS{PS}
\def\epsf@getbb#1{%
\openin\ps@stream=#1
\ifeof\ps@stream\ps@typeout{Error, File #1 not found}\else
   {\not@eoftrue \chardef\other=12
    \def\do##1{\catcode`##1=\other}\dospecials \catcode`\ =10
    \loop
       \if@psfile
	  \read\ps@stream to \epsf@fileline
       \else{
	  \obeyspaces
          \read\ps@stream to \epsf@tmp\global\let\epsf@fileline\epsf@tmp}
       \fi
       \ifeof\ps@stream\not@eoffalse\else
       \if@psfile\else
       \expandafter\epsf@test\epsf@fileline:. \\%
       \fi
          \expandafter\epsf@aux\epsf@fileline:. \\%
       \fi
   \ifnot@eof\repeat
   }\closein\ps@stream\fi}%
\long\def\epsf@test#1#2#3:#4\\{\def\epsf@testit{#1#2}
			\ifx\epsf@testit\epsf@start\else
\ps@typeout{Warning! File does not start with `\epsf@start'.  It may not be a PostScript file.}
			\fi
			\@psfiletrue} 
{\catcode`\%=12\global\let\epsf@percent=
\long\def\epsf@aux#1#2:#3\\{\ifx#1\epsf@percent
   \def\epsf@testit{#2}\ifx\epsf@testit\epsf@bblit
	\@atendfalse
        \epsf@atend #3 . \\%
	\if@atend	
	   \if@verbose{
		\ps@typeout{psfig: found `(atend)'; continuing search}
	   }\fi
        \else
        \epsf@grab #3 . . . \\%
        \not@eoffalse
        \global\no@bbfalse
        \fi
   \fi\fi}%
\def\epsf@grab #1 #2 #3 #4 #5\\{%
   \global\def\epsf@llx{#1}\ifx\epsf@llx\empty
      \epsf@grab #2 #3 #4 #5 .\\\else
   \global\def\epsf@lly{#2}%
   \global\def\epsf@urx{#3}\global\def\epsf@ury{#4}\fi}%
\def\epsf@atendlit{(atend)} 
\def\epsf@atend #1 #2 #3\\{%
   \def\epsf@tmp{#1}\ifx\epsf@tmp\empty
      \epsf@atend #2 #3 .\\\else
   \ifx\epsf@tmp\epsf@atendlit\@atendtrue\fi\fi}

\chardef\psletter = 11 
\chardef\other = 12

\newif \ifdebug 
\newif\ifc@mpute 
\c@mputetrue 

\let\then = \relax
\def\r@dian{pt }
\let\r@dians = \r@dian
\let\dimensionless@nit = \r@dian
\let\dimensionless@nits = \dimensionless@nit
\def\internal@nit{sp }
\let\internal@nits = \internal@nit
\newif\ifstillc@nverging
\def \Mess@ge #1{\ifdebug \then \message {#1} \fi}

{ 
	\catcode `\@ = \psletter
	\gdef \nodimen {\expandafter \n@dimen \the \dimen}
	\gdef \term #1 #2 #3%
	       {\edef \t@ {\the #1}
		\edef \t@@ {\expandafter \n@dimen \the #2\r@dian}%
		\t@rm {\t@} {\t@@} {#3}%
	       }
	\gdef \t@rm #1 #2 #3%
	       {{%
		\count 0 = 0
		\dimen 0 = 1 \dimensionless@nit
		\dimen 2 = #2\relax
		\Mess@ge {Calculating term #1 of \nodimen 2}%
		\loop
		\ifnum	\count 0 < #1
		\then	\advance \count 0 by 1
			\Mess@ge {Iteration \the \count 0 \space}%
			\Multiply \dimen 0 by {\dimen 2}%
			\Mess@ge {After multiplication, term = \nodimen 0}%
			\Divide \dimen 0 by {\count 0}%
			\Mess@ge {After division, term = \nodimen 0}%
		\repeat
		\Mess@ge {Final value for term #1 of 
				\nodimen 2 \space is \nodimen 0}%
		\xdef \Term {#3 = \nodimen 0 \r@dians}%
		\aftergroup \Term
	       }}
	\catcode `\p = \other
	\catcode `\t = \other
	\gdef \n@dimen #1pt{#1} 
}

\def \Divide #1by #2{\divide #1 by #2} 

\def \Multiply #1by #2
       {{
	\count 0 = #1\relax
	\count 2 = #2\relax
	\count 4 = 65536
	\Mess@ge {Before scaling, count 0 = \the \count 0 \space and
			count 2 = \the \count 2}%
	\ifnum	\count 0 > 32767 
	\then	\divide \count 0 by 4
		\divide \count 4 by 4
	\else	\ifnum	\count 0 < -32767
		\then	\divide \count 0 by 4
			\divide \count 4 by 4
		\else
		\fi
	\fi
	\ifnum	\count 2 > 32767 
	\then	\divide \count 2 by 4
		\divide \count 4 by 4
	\else	\ifnum	\count 2 < -32767
		\then	\divide \count 2 by 4
			\divide \count 4 by 4
		\else
		\fi
	\fi
	\multiply \count 0 by \count 2
	\divide \count 0 by \count 4
	\xdef \product {#1 = \the \count 0 \internal@nits}%
	\aftergroup \product
       }}

\def\r@duce{\ifdim\dimen0 > 90\r@dian \then   
		\multiply\dimen0 by -1
		\advance\dimen0 by 180\r@dian
		\r@duce
	    \else \ifdim\dimen0 < -90\r@dian \then  
		\advance\dimen0 by 360\r@dian
		\r@duce
		\fi
	    \fi}

\def\Sine#1%
       {{%
	\dimen 0 = #1 \r@dian
	\r@duce
	\ifdim\dimen0 = -90\r@dian \then
	   \dimen4 = -1\r@dian
	   \c@mputefalse
	\fi
	\ifdim\dimen0 = 90\r@dian \then
	   \dimen4 = 1\r@dian
	   \c@mputefalse
	\fi
	\ifdim\dimen0 = 0\r@dian \then
	   \dimen4 = 0\r@dian
	   \c@mputefalse
	\fi
	\ifc@mpute \then
		\divide\dimen0 by 180
		\dimen0=3.141592654\dimen0
		\dimen 2 = 3.1415926535897963\r@dian 
		\divide\dimen 2 by 2 
		\Mess@ge {Sin: calculating Sin of \nodimen 0}%
		\count 0 = 1 
		\dimen 2 = 1 \r@dian 
		\dimen 4 = 0 \r@dian 
		\loop
			\ifnum	\dimen 2 = 0 
			\then	\stillc@nvergingfalse 
			\else	\stillc@nvergingtrue
			\fi
			\ifstillc@nverging 
			\then	\term {\count 0} {\dimen 0} {\dimen 2}%
				\advance \count 0 by 2
				\count 2 = \count 0
				\divide \count 2 by 2
				\ifodd	\count 2 
				\then	\advance \dimen 4 by \dimen 2
				\else	\advance \dimen 4 by -\dimen 2
				\fi
		\repeat
	\fi		
			\xdef \sine {\nodimen 4}%
       }}

\def\Cosine#1{\ifx\sine\UnDefined\edef\Savesine{\relax}\else
		             \edef\Savesine{\sine}\fi
	{\dimen0=#1\r@dian\advance\dimen0 by 90\r@dian
	 \Sine{\nodimen 0}
	 \xdef\cosine{\sine}
	 \xdef\sine{\Savesine}}}	      

\def\psdraft{
	\def\@psdraft{0}
}
\def\psfull{
	\def\@psdraft{100}
}

\psfull

\newif\if@scalefirst
\def\psscalefirst{\@scalefirsttrue}
\def\psrotatefirst{\@scalefirstfalse}
\psrotatefirst

\newif\if@draftbox
\def\psnodraftbox{
	\@draftboxfalse
}
\def\psdraftbox{
	\@draftboxtrue
}
\@draftboxtrue

\newif\if@prologfile
\newif\if@postlogfile
\def\pssilent{
	\@noisyfalse
}
\def\psnoisy{
	\@noisytrue
}
\psnoisy
\newif\if@bbllx
\newif\if@bblly
\newif\if@bburx
\newif\if@bbury
\newif\if@height
\newif\if@width
\newif\if@rheight
\newif\if@rwidth
\newif\if@angle
\newif\if@clip
\newif\if@verbose
\def\@p@@sclip#1{\@cliptrue}

\newif\if@decmpr

\def\@p@@sfigure#1{\def\@p@sfile{null}\def\@p@sbbfile{null}
	        \openin1=#1.bb
		\ifeof1\closein1
	        	\openin1=\figurepath#1.bb
			\ifeof1\closein1
			        \openin1=#1
				\ifeof1\closein1%
				       \openin1=\figurepath#1
					\ifeof1
					   \ps@typeout{Error, File #1 not found}
						\if@bbllx\if@bblly
				   		\if@bburx\if@bbury
			      				\def\@p@sfile{#1}%
			      				\def\@p@sbbfile{#1}%
							\@decmprfalse
				  	   	\fi\fi\fi\fi
					\else\closein1
				    		\def\@p@sfile{\figurepath#1}%
				    		\def\@p@sbbfile{\figurepath#1}%
						\@decmprfalse
	                       		\fi%
			 	\else\closein1%
					\def\@p@sfile{#1}
					\def\@p@sbbfile{#1}
					\@decmprfalse
			 	\fi
			\else
				\def\@p@sfile{\figurepath#1}
				\def\@p@sbbfile{\figurepath#1.bb}
				\@decmprtrue
			\fi
		\else
			\def\@p@sfile{#1}
			\def\@p@sbbfile{#1.bb}
			\@decmprtrue
		\fi}

\def\@p@@sfile#1{\@p@@sfigure{#1}}

\def\@p@@sbbllx#1{
		\@bbllxtrue
		\dimen100=#1
		\edef\@p@sbbllx{\number\dimen100}
}
\def\@p@@sbblly#1{
		\@bbllytrue
		\dimen100=#1
		\edef\@p@sbblly{\number\dimen100}
}
\def\@p@@sbburx#1{
		\@bburxtrue
		\dimen100=#1
		\edef\@p@sbburx{\number\dimen100}
}
\def\@p@@sbbury#1{
		\@bburytrue
		\dimen100=#1
		\edef\@p@sbbury{\number\dimen100}
}
\def\@p@@sheight#1{
		\@heighttrue
		\dimen100=#1
   		\edef\@p@sheight{\number\dimen100}
}
\def\@p@@swidth#1{
		\@widthtrue
		\dimen100=#1
		\edef\@p@swidth{\number\dimen100}
}
\def\@p@@srheight#1{
		\@rheighttrue
		\dimen100=#1
		\edef\@p@srheight{\number\dimen100}
}
\def\@p@@srwidth#1{
		\@rwidthtrue
		\dimen100=#1
		\edef\@p@srwidth{\number\dimen100}
}
\def\@p@@sangle#1{
		\@angletrue
		\edef\@p@sangle{#1} 
}
\def\@p@@ssilent#1{ 
		\@verbosefalse
}
\def\@p@@sprolog#1{\@prologfiletrue\def\@prologfileval{#1}}
\def\@p@@spostlog#1{\@postlogfiletrue\def\@postlogfileval{#1}}
\def\@cs@name#1{\csname #1\endcsname}
\def\@setparms#1=#2,{\@cs@name{@p@@s#1}{#2}}
\def\ps@init@parms{
		\@bbllxfalse \@bbllyfalse
		\@bburxfalse \@bburyfalse
		\@heightfalse \@widthfalse
		\@rheightfalse \@rwidthfalse
		\def\@p@sbbllx{}\def\@p@sbblly{}
		\def\@p@sbburx{}\def\@p@sbbury{}
		\def\@p@sheight{}\def\@p@swidth{}
		\def\@p@srheight{}\def\@p@srwidth{}
		\def\@p@sangle{0}
		\def\@p@sfile{} \def\@p@sbbfile{}
		\def\@p@scost{10}
		\def\@sc{}
		\@prologfilefalse
		\@postlogfilefalse
		\@clipfalse
		\if@noisy
			\@verbosetrue
		\else
			\@verbosefalse
		\fi
}
\def\parse@ps@parms#1{
	 	\@psdo\@psfiga:=#1\do
		   {\expandafter\@setparms\@psfiga,}}
\newif\ifno@bb
\def\bb@missing{
	\if@verbose{
		\ps@typeout{psfig: searching \@p@sbbfile \space  for bounding box}
	}\fi
	\no@bbtrue
	\epsf@getbb{\@p@sbbfile}
        \ifno@bb \else \bb@cull\epsf@llx\epsf@lly\epsf@urx\epsf@ury\fi
}	
\def\bb@cull#1#2#3#4{
	\dimen100=#1 bp\edef\@p@sbbllx{\number\dimen100}
	\dimen100=#2 bp\edef\@p@sbblly{\number\dimen100}
	\dimen100=#3 bp\edef\@p@sbburx{\number\dimen100}
	\dimen100=#4 bp\edef\@p@sbbury{\number\dimen100}
	\no@bbfalse
}
\newdimen\p@intvaluex
\newdimen\p@intvaluey
\def\rotate@#1#2{{\dimen0=#1 sp\dimen1=#2 sp
		  \global\p@intvaluex=\cosine\dimen0
		  \dimen3=\sine\dimen1
		  \global\advance\p@intvaluex by -\dimen3
		  \global\p@intvaluey=\sine\dimen0
		  \dimen3=\cosine\dimen1
		  \global\advance\p@intvaluey by \dimen3
		  }}
\def\compute@bb{
		\no@bbfalse
		\if@bbllx \else \no@bbtrue \fi
		\if@bblly \else \no@bbtrue \fi
		\if@bburx \else \no@bbtrue \fi
		\if@bbury \else \no@bbtrue \fi
		\ifno@bb \bb@missing \fi
		\ifno@bb \ps@typeout{FATAL ERROR: no bb supplied or found}
			\no-bb-error
		\fi
		\count203=\@p@sbburx
		\count204=\@p@sbbury
		\advance\count203 by -\@p@sbbllx
		\advance\count204 by -\@p@sbblly
		\edef\ps@bbw{\number\count203}
		\edef\ps@bbh{\number\count204}
		\if@angle 
			\Sine{\@p@sangle}\Cosine{\@p@sangle}
	        	{\dimen100=\maxdimen\xdef\r@p@sbbllx{\number\dimen100}
					    \xdef\r@p@sbblly{\number\dimen100}
			                    \xdef\r@p@sbburx{-\number\dimen100}
					    \xdef\r@p@sbbury{-\number\dimen100}}
                        \def\minmaxtest{
			   \ifnum\number\p@intvaluex<\r@p@sbbllx
			      \xdef\r@p@sbbllx{\number\p@intvaluex}\fi
			   \ifnum\number\p@intvaluex>\r@p@sbburx
			      \xdef\r@p@sbburx{\number\p@intvaluex}\fi
			   \ifnum\number\p@intvaluey<\r@p@sbblly
			      \xdef\r@p@sbblly{\number\p@intvaluey}\fi
			   \ifnum\number\p@intvaluey>\r@p@sbbury
			      \xdef\r@p@sbbury{\number\p@intvaluey}\fi
			   }
			\rotate@{\@p@sbbllx}{\@p@sbblly}
			\minmaxtest
			\rotate@{\@p@sbbllx}{\@p@sbbury}
			\minmaxtest
			\rotate@{\@p@sbburx}{\@p@sbblly}
			\minmaxtest
			\rotate@{\@p@sbburx}{\@p@sbbury}
			\minmaxtest
			\edef\@p@sbbllx{\r@p@sbbllx}\edef\@p@sbblly{\r@p@sbblly}
			\edef\@p@sbburx{\r@p@sbburx}\edef\@p@sbbury{\r@p@sbbury}
		\fi
		\count203=\@p@sbburx
		\count204=\@p@sbbury
		\advance\count203 by -\@p@sbbllx
		\advance\count204 by -\@p@sbblly
		\edef\@bbw{\number\count203}
		\edef\@bbh{\number\count204}
}
\def\in@hundreds#1#2#3{\count240=#2 \count241=#3
		     \count100=\count240	
		     \divide\count100 by \count241
		     \count101=\count100
		     \multiply\count101 by \count241
		     \advance\count240 by -\count101
		     \multiply\count240 by 10
		     \count101=\count240	
		     \divide\count101 by \count241
		     \count102=\count101
		     \multiply\count102 by \count241
		     \advance\count240 by -\count102
		     \multiply\count240 by 10
		     \count102=\count240	
		     \divide\count102 by \count241
		     \count200=#1\count205=0
		     \count201=\count200
			\multiply\count201 by \count100
		 	\advance\count205 by \count201
		     \count201=\count200
			\divide\count201 by 10
			\multiply\count201 by \count101
			\advance\count205 by \count201
		     \count201=\count200
			\divide\count201 by 100
			\multiply\count201 by \count102
			\advance\count205 by \count201
		     \edef\@result{\number\count205}
}
\def\compute@wfromh{
		\in@hundreds{\@p@sheight}{\@bbw}{\@bbh}
		\edef\@p@swidth{\@result}
}
\def\compute@hfromw{
	        \in@hundreds{\@p@swidth}{\@bbh}{\@bbw}
		\edef\@p@sheight{\@result}
}
\def\compute@handw{
		\if@height 
			\if@width
			\else
				\compute@wfromh
			\fi
		\else 
			\if@width
				\compute@hfromw
			\else
				\edef\@p@sheight{\@bbh}
				\edef\@p@swidth{\@bbw}
			\fi
		\fi
}
\def\compute@resv{
		\if@rheight \else \edef\@p@srheight{\@p@sheight} \fi
		\if@rwidth \else \edef\@p@srwidth{\@p@swidth} \fi
}
\def\compute@sizes{
	\compute@bb
	\if@scalefirst\if@angle
	\if@width
	   \in@hundreds{\@p@swidth}{\@bbw}{\ps@bbw}
	   \edef\@p@swidth{\@result}
	\fi
	\if@height
	   \in@hundreds{\@p@sheight}{\@bbh}{\ps@bbh}
	   \edef\@p@sheight{\@result}
	\fi
	\fi\fi
	\compute@handw
	\compute@resv}

\def\psfig#1{\vbox {
	\ps@init@parms
	\parse@ps@parms{#1}
	\compute@sizes
	\ifnum\@p@scost<\@psdraft{
		\special{ps::[begin] 	\@p@swidth \space \@p@sheight \space
				\@p@sbbllx \space \@p@sbblly \space
				\@p@sbburx \space \@p@sbbury \space
				startTexFig \space }
		\if@angle
			\special {ps:: \@p@sangle \space rotate \space} 
		\fi
		\if@clip{
			\if@verbose{
				\ps@typeout{(clip)}
			}\fi
			\special{ps:: doclip \space }
		}\fi
		\if@prologfile
		    \special{ps: plotfile \@prologfileval \space } \fi
		\if@decmpr{
			\if@verbose{
				\ps@typeout{psfig: including \@p@sfile.Z \space }
			}\fi
			\special{ps: plotfile "`zcat \@p@sfile.Z" \space }
		}\else{
			\if@verbose{
				\ps@typeout{psfig: including \@p@sfile \space }
			}\fi
			\special{ps: plotfile \@p@sfile \space }
		}\fi
		\if@postlogfile
		    \special{ps: plotfile \@postlogfileval \space } \fi
		\special{ps::[end] endTexFig \space }
		\vbox to \@p@srheight sp{
			\hbox to \@p@srwidth sp{
				\hss
			}
		\vss
		}
	}\else{
		\if@draftbox{		
			\hbox{\frame{\vbox to \@p@srheight sp{
			\vss
			\hbox to \@p@srwidth sp{ \hss \@p@sfile \hss }
			\vss
			}}}
		}\else{
			\vbox to \@p@srheight sp{
			\vss
			\hbox to \@p@srwidth sp{\hss}
			\vss
			}
		}\fi

	}\fi
}}
\psfigRestoreAt
\let\@=\LaTeXAtSign